\documentclass[11pt]{article}
\usepackage{amssymb}
\usepackage{amsmath}
\usepackage{amsfonts}
\usepackage{bm}
\usepackage{epsfig}
\usepackage{graphicx}

\newcounter{myfig}

\title{\vskip65pt
\bf Hilbert fluid dynamics equations expressed in Chapman-Enskog pressure
tensor and heat current}
\author{\\[-0.4cm]
{\bf Soderholm, Lars H.}\\
Mekanik, KTH, SE-10044 Stockholm, Sweden\\
\texttt{lars.soderholm@mech.kth.se}\\
\\[-5mm]
}
\date{ }

\begin{document}
\maketitle

\section{Abstract}
The connection between the Chapman-Enskog and Hilbert expansions is investigated in detail. In particular
the fluid dynamics equations of any order in the Hilbert expansion are given in terms of the pressure tensor
and heat current of the Chapman-Enskog expansion
\vskip 10pt
\noindent \emph{Key words:} Hilbert expansion; Chapman-Enskog expansion.

\section{Introduction}
The most convenient way to calculate heat current and pressure tensor is
that of the Chapman-Enskog method, see Grad \cite{Grad HdB}. In particular, 
the terms up to second and
higher order have been calculated and at least to second order there is an
agreed convention of notation for the terms. But in many occasions the
Hilbert method is preferred because of the difficulties with the Burnett
equations, see Sone \cite{Sonebok}. Already in the Handbuch article \cite{Grad HdB}, 
Grad made clear that the
Chapman-Enskog method and Hilbert method are two ways of expressing the same
thing. But Grad adds "to confirm this directly is an intricate task due to
the different formalisms". So, his proof is somewhat abstract. It is the
object of the present work to do the explicit confirmation that Grad does
not do. 

The benefit of this is that the fluid dynamics equations of the
Hilbert expansion are given explicitly in terms of the pressure tensor and
heat current of the Chapman-Enskog method. This means that results that have
been derived by the Chapman-Enskog method directly can be used in the
Hilbert expansion. In the interesting paper \cite{Chekmarevs} by Chekmarev and Chekmareva
multiple scale methods are applied in the Hilbert expansion. The results
of the present paper should considerably simplify and generalize 
their results to second order. In the paper \cite{Soderholm RGD25}
the results by Chekmarev and Chekmareva were generalized but derived
from the Burnett equations. This could be considered unsatisfactory
as the Burnett equations are mathematically unstable. But here we have
been able to show that this result was nevertheless correct. We 
have shown that the fluid dynamics equations of the
Hilbert expansion in fact can be obtained from the corresponding equations of
the Chapman-Enskog expansion by expanding the fluid dynamics fields.

\section{Chapman-Enskog expansion and expansion of fluid dynamics variables}
\setcounter{equation}{0}
We write the Boltzmann equation in dimensionless variables ($\varepsilon $
is the Knudsen number)%
\[
\varepsilon \mathcal{D}f=Q(f,f),
\]%
where the streaming operator is ($\mathbf{F}$ is the force per mass)%
\[
\mathcal{D=}\frac{\partial }{\partial t}+\mathbf{c\cdot }\frac{\partial }{%
\partial \mathbf{x}}+\mathbf{F\cdot }\frac{\partial }{\partial \mathbf{c}}.
\]
See Cercignani \cite{Cercignani bok}The fluid dynamic fields are given by%
\begin{eqnarray}
m\int fd^{3}c &=&\rho ,  \nonumber \\
m\int f\mathbf{c}d^{3}c &=&\rho \mathbf{v=j,}  \label{fluid dynamics fields}
\\
\frac{m}{2}\int fc^{2}d^{3}c &=&\rho (\frac{v^{2}}{2}+\frac{3k_{B}T}{2m})=e 
\nonumber
\end{eqnarray}%
In the Chapman-Enskog expansion is based on the important auxiliary
condition, that only the zero order distribution function contributes to the
flud dynamics fields. Hence ({\ref{fluid dynamics fields}}) is satisfied
also with $f$ replaced by $f_{CE0}.$We summarize the fluid dynamical fields 
\[
w=\left\{ 
\begin{array}{c}
\rho  \\ 
\mathbf{v} \\ 
T%
\end{array}%
\right\} .
\]%
The components of $w$ are written $w_{a}$ where $a=1...5$. The
Chapman-Enskog series is written ($\varepsilon $ is the Knudsen number)%
\[
f_{CE}=f_{CE0}(w,\mathbf{c})+\varepsilon f_{CE1}(w,\mathbf{c})+...
\]%
Here, $f_{CEr}(w,\mathbf{c})$ stands for a function of $w$ and its $r$ first
derivatives and of $\mathbf{c}$, the molecular velocity. In the sequel we
don't write out $\mathbf{c}$ explicitly.

Let us now assume that the fluid dynamics variables can be expanded in the
Knudsen number 
\[
w=w_{0}+\varepsilon w_{1}+...
\]%
Applying this expansion in $f_{CEr}(w,\mathbf{c})$ we find 
\[
f_{CEr}(w)=f_{CEr,0}(w_{0})+\varepsilon f_{CEr,1}(w_{0},w_{1})+...
\]%
Here $f_{CEr,s}$ is a function of $w_{0},w_{1},...w_{s}$ and their $r$ first
derivatives. 

\bigskip We first of all consider the zero order term $f_{CE0}.$ When we
expand it we obtain to lowest order $f_{CE0,0}(w_{0})$. This is a Maxwellian
for the fluid dynamics fields $w_{0}$. In other words, this is the zero
order term in the Hilbert expansion%
\[
f_{CE0,0}=f_{H0}.
\]

Let me write the sum of the first $r$ terms in the Chapman-Enskog expansion
and introduce this expansion of the fluid dynamics fields%
\begin{eqnarray*}
&&f_{CE0}+\varepsilon f_{CE1}+...\varepsilon ^{n}f_{CEr} \\
&=&f_{CE0,0}+\varepsilon (f_{CE0,1}+f_{CE1,0})+...\varepsilon
^{r}\sum\limits_{s=0}^{r}f_{CEr-s,s}+O(\varepsilon ^{r+1})
\end{eqnarray*}%
Thus we obtain%
\begin{eqnarray*}
\rho  =m\int f_{CE0}d^{3}c=m\int (f_{CE0}+\varepsilon
f_{CE1}+...\varepsilon ^{r}f_{CEr})d^{3}c \\
=m\int f_{CE0,0}d^{3}c+\varepsilon m\int
(f_{CE0,1}+f_{CE1,0})d^{3}c+...\\ 
+\varepsilon ^{r}m\int
\sum\limits_{s=0}^{r}f_{CEr-s,s}d^{3}c+O(\varepsilon ^{s+1})
\end{eqnarray*}%
Identifying the terms we find%
\begin{equation}
\rho _{r}=m\int \sum\limits_{s=0}^{r}f_{CEr-s,s}d^{3}c.  \label{n som serie}
\end{equation}

In the same way we obtain 
\begin{equation}
\mathbf{j}_{r}=(\rho \mathbf{v)}_{r}=\sum\limits_{s=0}^{r}\rho _{r-s}%
\mathbf{v}_{s}=m\int \sum\limits_{s=0}^{r}f_{CEr-s,s}\mathbf{c}d^{3}c.
\label{nv som serie}
\end{equation}%
Finally we have%
\begin{eqnarray}
e_{r} &=&[\rho (\frac{v^{2}}{2}+\frac{3k_{B}T}{2m})]_{r}  \nonumber \\
&=&\sum\limits_{s=0}^{r}\rho _{r-s}(\sum\limits_{p=0}^{s}\frac{1}{2}%
\mathbf{v}_{s-p}\mathbf{\cdot v}_{p}+\frac{3k_{B}}{2m}T_{s})  \nonumber \\
&=&\frac{m}{2}\int \sum\limits_{s=0}^{r}f_{CEr-s,s}c^{2}d^{3}c
\label{n(v2 + T) som serie}
\end{eqnarray}%
From {(\ref{n som serie}-\ref{n(v2 + T) som serie}) }we conclude that if we
assume that the fluid dynamics field are expanded in a series in the Knudsen
number, the corresponding terms to order $r$ of the fields are given as the
projections of 
\begin{equation}
\sum\limits_{s=0}^{r}f_{CEr-s,s}.  \label{Hilbertkandidat}
\end{equation}%
We shall see that $\sum\limits_{s=0}^{r}f_{CEr-s,s}$ is in fact the
coefficient $f_{Hr}$ of $\varepsilon ^{r}$ in the Hilbert expansion. 

\section{The Hilbert distribution function expressed in the Chapman-Enskog
distribution function}

Let us now consider the terms up to order $r$ in the Chapman-Enskog
expression of the collision term. We then expand the fluid dynamics fields
and obtain 
\begin{eqnarray*}
&&\sum\limits_{p+q=0}^{r}\varepsilon ^{p+q}Q(f_{CEp},f_{CEq}) \\
&=&\sum\limits_{p+q+i+j=0}^{r}\varepsilon
^{p+q+i+j}Q(f_{CEp,i},f_{CEq,j})+O(\varepsilon ^{r+1})
\end{eqnarray*}%
In the coefficient of $\varepsilon ^{r}$ we put $p+i=k,q+j=l$ so that it
becomes 
\begin{eqnarray*}
&&\sum\limits_{k+l=r}Q(f_{CEk-p,p},f_{CEl-q,q}) \\
&=&\sum\limits_{k+l=r}Q(\sum\limits_{p=0}^{k}f_{CEk-p,p},\sum%
\limits_{q=0}^{l}f_{CEl-q,q})
\end{eqnarray*}%
We obtains sums of the kind ({\ref{Hilbertkandidat}}).

If we do the corresponding thing in the streaming term we find that the
coefficient of  $\varepsilon ^{r-1}$is 
\[
\mathcal{D}\sum\limits_{p=0}^{r-1}f_{CEr-1-p,p}
\]
So from the Chapman-Enskog equations of order $0,1,...r$ we find that%
\begin{eqnarray}
&&Q(\sum\limits_{p=0}^{r}f_{CEr-p,p},f_{CE0,0})+Q(f_{CE0,0},\sum%
\limits_{p=0}^{r}f_{CEr-p,p})  \nonumber \\
&=&-\sum\limits_{k+l=r}Q(\sum\limits_{p=1}^{r-1}f_{CEk-p,p},\sum%
\limits_{q=1}^{r-1}f_{CEl-q,q})  \label{Hilbertekvation uttryckt i CE} \\
&&+\mathcal{D}\sum\limits_{p=0}^{r-1}f_{CEr-1-p,p}  \nonumber
\end{eqnarray}

Now we proceed by induction. We assume that the sum ({\ref{Hilbertkandidat}}%
) to order $r-1$ is the corresponding term in the Hilbert expansion 
\begin{equation}
f_{Hr-1}=\sum\limits_{s=0}^{r-1}f_{CEr-1-s,s}.  \label{samband Hilbert CE}
\end{equation}%
We can then write ({\ref{Hilbertekvation uttryckt i CE}}) as 
\begin{eqnarray*}
Q(\sum\limits_{p=0}^{r}f_{CEr-p,p},f_{H0})+Q(f_{H0},\sum%
\limits_{p=0}^{r}f_{CEr-p,p}) \\
=-\sum\limits_{s=1}^{r-1}Q(f_{Hr-s},f_{Hs}) +\mathcal{D}f_{Hr-1}
\end{eqnarray*}%
This means that 
\begin{equation}
\sum_{s=0}^{r}f_{CEr-s,s}  \label{Hilbertkandidat r}
\end{equation}%
satisfies the equation 
\begin{eqnarray}
Q(f_{Hr},f_{H0})+Q(f_{H0},f_{Hr})  \label{Hilbert ekvation ordning r} \\
=-\sum\limits_{s=1}^{r-1}Q(f_{Hr-s},f_{Hs}) +\mathcal{D}f_{Hr-1}  \nonumber
\end{eqnarray}%
for $f_{Hr}$ in the Hilbert expansion. We assume the existence of a solution
and its uniqueness up to a linear combination of $1,\mathbf{c,}c^{2}$. But
according to {(\ref{n som serie}-\ref{n(v2 + T) som serie}) the function (%
\ref{Hilbertkandidat r}) also has the correct projections for }$f_{Hr}$ {on
the functions }$1,\mathbf{c,}c^{2}.$ So by induction this is the function $%
f_{Hr}$ of the Hilbert expansion. We conclude that the simple relation
between the Chapman-Enskog and Hilbert distribution functions is 
\begin{equation}
f_{Hr}=\sum\limits_{s=0}^{r}f_{CEr-s,s}.  \label{hilbert fr slutresultat}
\end{equation}

\section{Conservation laws and the Chapman-Enskog equations}
Let us to start by writing down the conservation laws. They are 
\begin{eqnarray}
\rho _{,t}+\mathbf{\nabla \cdot j} &=&0  \nonumber \\
\mathbf{j}_{,t}+\mathbf{\nabla \cdot J} &\mathbf{=}&\rho \mathbf{F}
\label{conservation laws} \\
e_{,t}+\mathbf{\nabla \cdot j}_{e} &=&\mathbf{F}\cdot \mathbf{j}  \nonumber
\end{eqnarray}%
Here%
\begin{equation}
\mathbf{J=}m\int f\mathbf{cc}d^{3}c=\mathbf{P+}\rho \mathbf{vv=P+jv,}
\label{momentum current}
\end{equation}%
where%
\begin{equation}
\mathbf{P=}m\int f\mathbf{c}^{\prime }\mathbf{c}^{\prime }d^{3}c
\label{pressure tensor}
\end{equation}%
and 
\begin{eqnarray}
\mathbf{j}_{e} &=&\int \frac{mc^{2}}{2}f\mathbf{c}d^{3}c
\label{energy current} \\
&=&e\mathbf{v+q+P\cdot v,}  \nonumber
\end{eqnarray}%
\begin{equation}
\mathbf{q=}\frac{m}{2}\int c^{\prime 2}f\mathbf{c}^{\prime }d^{3}c\mathbf{.}
\label{heat current}
\end{equation}

If we instead calculate the moments from $f_{CEr}$ we obtain to zero order%
\begin{equation}
\mathbf{J}_{CE0}{=}m\int f_{CE0}\mathbf{cc}d^{3}c = \mathbf{P}_{CE0}%
\mathbf{+jv,}  \label{momentum current CE0}
\end{equation}%
\begin{equation}
\mathbf{P}_{CE0}\mathbf{=}m\int f_{CE0}\mathbf{c}^{\prime }\mathbf{c}%
^{\prime }d^{3}c=p\mathbf{1}  \label{pressure tensor CE0}
\end{equation}
\begin{eqnarray}
\mathbf{j}_{eCE0} &=&\int \frac{mc^{2}}{2}f_{CE0}\mathbf{c}d^{3}c
\label{energy current CE0} \\
&=&e\mathbf{v+q}_{CE0}\mathbf{+P}_{CE0}\mathbf{\cdot v,}  \nonumber
\end{eqnarray}%
\begin{equation}
\mathbf{q}_{CE0}\mathbf{=}\frac{m}{2}\int c^{\prime 2}f\mathbf{c}^{\prime
}d^{3}c=\mathbf{0.}  \label{heat current CE0}
\end{equation}%
For $r\geq 1$ we have%
\begin{equation}
\mathbf{J}_{CEr}\mathbf{=}m\int f_{CEr}\mathbf{cc}d^{3}c=\mathbf{P}_{CEr}%
\mathbf{,}  \label{momentum current CEr}
\end{equation}%
\begin{equation}
\mathbf{P}_{CEr}\mathbf{=}m\int f_{CEr}\mathbf{c}^{\prime }\mathbf{c}%
^{\prime }d^{3}c  \label{pressure tensor CEr}
\end{equation}
\begin{eqnarray}
\mathbf{j}_{eCEr} &=&\int \frac{mc^{2}}{2}f_{CEr}\mathbf{c}d^{3}c
\label{energy current CEr} \\
&=&\mathbf{q}_{CEr}\mathbf{+P}_{CEr}\mathbf{\cdot v,}  \nonumber
\end{eqnarray}%
\begin{equation}
\mathbf{q}_{CEr}\mathbf{=}\frac{m}{2}\int c^{\prime 2}f_{CEr}\mathbf{c}%
^{\prime }d^{3}c\mathbf{.}  \label{heat current CEr}
\end{equation}%
$\mathbf{P}_{CEr}$ and $\mathbf{q}_{CEr}$ and hence $\mathbf{J}_{CEr}$ and $%
\mathbf{j}_{eCEr}$ are functions of the fluid dynamics fields $w$ and its $r$
first derivatives. This means that we can expand the fluid dynamics fields
in them to obtain%
\[
\mathbf{P}_{CEr}=\mathbf{P}_{CEr,0}+\varepsilon \mathbf{P}_{CEr,1}+...,
\]%
where $\mathbf{P}_{CEr,s}$ is a function of $w_{0},w_{1},...w_{s}$ and its $r
$ first derivatives. 

The Chapman-Enskog equations to order $r$ are obtained if we make the
approximations 
\begin{eqnarray}
\mathbf{P} &\mathbf{\rightarrow }&\sum\limits_{s=0}^{r}\varepsilon ^{s}%
\mathbf{P}_{CEs},  \label{approx Pq CE} \\
\mathbf{q} &\mathbf{\rightarrow }&\sum\limits_{s=0}^{r}\varepsilon ^{s}%
\mathbf{q}_{CEs}  \nonumber
\end{eqnarray}%
in the expressions for $\mathbf{J}$ and $\mathbf{j}_{e}$ and substitute into
the equations of balance. The equations are then%
\begin{eqnarray}
\rho _{,t}+\mathbf{\nabla \cdot j} &=&0  \nonumber \\
\mathbf{j}_{,t}+\mathbf{\nabla \cdot (}\sum\limits_{s=0}^{r}\varepsilon ^{s}%
\mathbf{\mathbf{J}}_{CEs}\mathbf{)} &\mathbf{=}&\rho \mathbf{F}
\label{CE-ekvationerna} \\
e_{,t}+\mathbf{\nabla \cdot (}\sum\limits_{s=0}^{r}\varepsilon ^{s}\mathbf{j%
}_{eCEs}\mathbf{)} &=&\mathbf{F}\cdot \mathbf{j}  \nonumber
\end{eqnarray}

\section{Integrability conditions for the Hilbert expansion}\label{sec math form}

We consider the equation {(\ref{Hilbert ekvation ordning r}) }of the Hilbert
expansion where we now replace $r$ by $r+1$. The condition of integrability
is that is 
\begin{equation}
\int \left\{ 
\begin{array}{c}
m \\ 
m\mathbf{c} \\ 
mc^{2}/2%
\end{array}%
\right\} \mathcal{D}f_{Hr}d^{3}c=0.  \label{integrability cond Hr}
\end{equation}%
We now use the relation {(\ref{hilbert fr slutresultat}) to express this
condition in terms of the Chapman-Enskog distribution function. To that end
we need the corresponding integrals f\"{o}r the Chapman-Enskog terms. }To
zero order we obtain%
\begin{eqnarray}
m\int \mathcal{D}f_{CE0}d^{3}c &=&\rho _{,t}+\mathbf{\nabla \cdot j,} 
\nonumber \\
m\int \mathcal{D}f_{CEr}\mathbf{c}d^{3}c &=&\mathbf{j}_{,t}+\mathbf{\nabla
\cdot J}_{CE0}\mathbf{-}\rho \mathbf{F,}  \label{proj Df CE0} \\
m\int \frac{c^{2}}{2}\mathcal{D}f_{CE0}\mathbf{c}d^{3}c &=&e_{,t}+\mathbf{%
\nabla \cdot j}_{eCE0}-\mathbf{F}\cdot \mathbf{j.}  \nonumber
\end{eqnarray}%
To order $r\geq 1$ there will be no contributions to the time derivatives 
\begin{eqnarray}
m\int \mathcal{D}f_{CEr}d^{3}c &=&0,  \nonumber \\
m\int \mathcal{D}f_{CEr}\mathbf{c}d^{3}c &=&\mathbf{\nabla \cdot J}_{CEr},
\label{proj Df CEr} \\
m\int \frac{c^{2}}{2}\mathcal{D}f_{CEr}\mathbf{c}d^{3}c &=&\mathbf{\nabla
\cdot j}_{eCEr}.  \nonumber
\end{eqnarray}

If we now expand the fluid dynamics fields we obtain%
\begin{eqnarray}
m\int \mathcal{D}f_{CE0,s}d^{3}c &=&\rho _{s,t}+\mathbf{\nabla \cdot j}_{s},
\nonumber \\
m\int \mathcal{D}f_{CE0,s}\mathbf{c}d^{3}c &=&\mathbf{j}_{s,t}+\mathbf{%
\nabla \cdot J}_{CE0,s}\mathbf{-}\rho _{s}\mathbf{F,}  \label{proj Df CE0,s}
\\
m\int \frac{c^{2}}{2}\mathcal{D}f_{CE0,s}\mathbf{c}d^{3}c &=&e_{s,t}+\mathbf{%
\nabla \cdot j}_{eCE0,s}-\mathbf{F}\cdot \mathbf{j}_{s}.  \nonumber
\end{eqnarray}%
For $r\geq 1$ we have%
\begin{eqnarray}
m\int \mathcal{D}f_{CEr,s}d^{3}c &=&0,  \nonumber \\
m\int \mathcal{D}f_{CEr,s}\mathbf{c}d^{3}c &=&\mathbf{\nabla \cdot J}%
_{CEr,s},  \label{proj Df CEr,s} \\
m\int \frac{c^{2}}{2}\mathcal{D}f_{CEr}\mathbf{c}d^{3}c &=&\mathbf{\nabla
\cdot j}_{eCEr,s}.  \nonumber
\end{eqnarray}%
Now, using the expression for $f_{Hr}$ in terms of $f_{CEr-s,s}$ we finally
obtain the conditions of integrability to order $r$%
\begin{eqnarray}
\rho _{r,t}+\mathbf{\nabla \cdot j}_{r} &=&0,  \nonumber \\
\mathbf{j}_{r,t}+\mathbf{\nabla \cdot }\sum\limits_{s=0}^{r}\mathbf{J}%
_{CEr-s,s}\mathbf{-}\rho _{r}\mathbf{F}\mathbf{=0,} &&
\label{integrability condition Hr result} \\
e_{r,t}+\mathbf{\nabla \cdot }\sum\limits_{s=0}^{r}\mathbf{j}_{eCE0r-s,s}-%
\mathbf{F}\cdot \mathbf{j}_{r} &=&0.  \nonumber
\end{eqnarray}%
They are the fluid dynamics equations of the Hilbert expansion. Note that
the same equations are obtained if we start with the Chapman-Enskog
equations of order $r$, expand the fluid dynamics fields and pick out the
coefficient of $\varepsilon ^{r}$. 

Let us now as an example\ consider the linearized Boltzmann equation and
find the Hilbert fluid dynamics equations to second order. We have, see
Chapman \&\ Cowling 
\begin{eqnarray}
\mathbf{P}_{CE1} &=&-2\mu <\mathbf{\nabla v}>,  \nonumber \\
\mathbf{P}_{CE2} &=&\frac{\mu ^{2}}{\rho T}(\varpi _{3}-\varpi _{2})\langle 
\mathbf{\nabla \nabla }T\rangle -\varpi _{2}\frac{\mu ^{2}}{\rho ^{2}}%
\langle \mathbf{\nabla \nabla }\rho \rangle ,  \label{upp till Burnett P,q}
\\
\mathbf{q}_{CE1} &=&-\kappa \,\mathbf{\nabla }\,\,T,  \nonumber \\
\mathbf{q}_{CE2} &=&\frac{\mu ^{2}}{\rho }[(\frac{\theta _{4}}{2})\triangle 
\mathbf{v}+(\frac{\theta _{4}}{6}-\frac{2\theta _{2}}{3})\mathbf{\nabla
(\nabla \cdot v)].}  \nonumber
\end{eqnarray}%
Hence, the Hilbert fluid dynamics equations to second order are%
\begin{eqnarray}
\rho _{2,t}+\rho _{0}\mathbf{\nabla \cdot v}_{2} =0,  \nonumber \\
\rho _{0}\mathbf{v}_{2,t} =-\mathbf{\nabla }p_{2}+2\mu _{0}(\triangle 
\mathbf{v}_{1}+\frac{2}{3}\mathbf{\nabla (\nabla \cdot v}_{1}\mathbf{)})
\label{Hilberts ordning 2} \\
-[\frac{2\mu ^{2}}{3\rho T}(\varpi _{3}-\varpi _{2})]_{0}\triangle \mathbf{%
\nabla }T_{0}+[\varpi _{2}\frac{2\mu ^{2}}{3\rho ^{2}}]_{0}\triangle \mathbf{%
\nabla }\rho _{0}, \\
\frac{3k_{B}\rho _{0}}{2m}(T_{2,t}+T_{0}\mathbf{\nabla \cdot v}_{2})
=\kappa _{0}\triangle T_{1}-[\frac{2\mu ^{2}}{3\rho }(\theta _{4}-\theta
_{2})]_{0}\triangle \mathbf{(\nabla \cdot v}_{0}\mathbf{)}.  \nonumber
\end{eqnarray}

\section*{Acknowledgements}
Discussions with A.V. Bobylev and Y. Sone have been stimulating.

\end{document}